IN DEFENCE OF CAMPBELL'S THEOREM AS A FRAME FOR NEW PHYSICS

Paul S. Wesson

Departments of Physics and Applied Mathematics

University of Waterloo, Waterloo, Ontario N2L 3G1, Canada

("People who live in glass houses should not throw stones" – Old English saying)

PACs:  04.50+h, 04.20.Ex, 04.20.Cv

Abstract:   The Campbell-Magaard theorem is widely seen as a way of embedding Einstein's 4D theory of general relativity in a 5D theory of the Kaluza-Klein type.  We give a brief history of the theorem, present a short account of it, and show that it provides a geometrical frame for new physics related to the unification of the forces.  Anderson's recent vituperative attack on Campbell's theorem (gr-qc/0409122) is errant.

Addresses:  Mail to address above, phone (519) 888-4567 Ext. 2215, fax (519) 746-8115.

1. <u>Introduction</u>

Campbell's theorem, as developed by Magaard and others, has in recent years come to be regarded as a geometrical justification for extending Einstein's theory of general relativity to more than the 4 dimensions of spacetime with which we are familiar [1-4]. Anderson, however, has newly attacked the CM theorem, and argued that it is inadequate and inappropriate as a protective theorem for relativistic field equations [5]. We will here show that, on the contrary, the majority opinion has veracity. We will also outline some physical problems which might repay the efforts of an unblinkered investigator.

That the properties of N-dimensional Riemannian spaces have relevance to physics is apparent from even a casual perusal of the modern literature. There one finds work particularly on $N = 5, 10, 11$ and 26. This pertains to Kaluza-Klein theory, supersymmetry, supergravity and string theory. The approaches may be different, but have in common the goal of unifying gravity (as described by $N = 4$ general relativity) with the interactions of particle physics. The case $N = 5$ has recently undergone a transformation, in that the "cylinder" condition of old Kaluza-Klein theory has been dropped, to yield non-compactified gravity. This exists in two main versions. Induced-matter theory uses an unconstrained extra dimension to account for the origin and nature of energy; while membrane theory uses a singular surface in 5D (the "brane") to localize particle forces, this separating regions (the "bulk") wherein operates gravity. There are many results known for both of these versions of 5D gravity. Extensive bibliographies are to be found in references [3], [4] and [5], so below we will quote only key sources. 5D gravity is widely regarded as the low-energy limit of even higher-dimensional theories, so it is not surprising that many workers have considered exact solutions of the 5D field equations. These in terms of the Ricci tensor are frequently taken to be $R_{AB} = 0$ (A, B = 0, 123, 4). By contrast, the field equations of 4D general relativity, in terms of the Einstein tensor and the energy-momentum tensor, can be taken to be $G_{\alpha\beta} = T_{\alpha\beta}$ $(\alpha, \beta = 0, 123)$. These may be augmented by the addition of a term involving the cosmological "constant" $\Lambda$, but it is convenient instead to include this term in the right-hand side of Einstein's equations, where it represents a (possibly variable) energy density for the vacuum.

From the viewpoint of physics, as opposed to differential geometry, the CM theorem has practical importance because it indicates that the 15 field equations $R_{AB} = 0$ <u>contain</u> the 10 field equations $G_{\alpha\beta} = T_{\alpha\beta}$ [3,4]. This is not only a result of considerable algebraic elegance, but also has philosophical implications, insofar as one can (if so desired) regard geometry as the seat of matter. (The other 5 field equations can be cast as a set of 4 conservation equations and a scalar wave equation: see for example ref. 4, pp. 58-66 and below.) There are other, more mathematical formulations of the CM theorem, and we will quote and give an heuristic proof of one such below. But as regards modern physics, interest in the CM theorem revolves largely around the fact that the 5D field equations provide an embedding for the 4D ones.

2. <u>A Short History of the Theorem</u>

John Edward Campbell was a professor of mathematics at Oxford whose book "A Course of Differential Geometry" was published posthumously in 1926 [1]. The book is basically a set of lecture notes on the algebraic properties of ND Riemannian manifolds, and the question of embeddings is treated in the latter part (notably chapters 12 and 14). However, what is nowadays called Campbell's theorem is there only sketched. The author had intended to add a chapter dealing with the relation between abstract spaces and Einstein's theory of general relativity (which was then a recent addition to physics), but died before he could complete it.

The book was compiled with the aid of Campbell's colleague, E.B. Elliott, but while accurate is certainly incomplete.

The problem of embedding an ND (pseudo-) Riemannian manifold in a Ricci-flat space of one higher dimension was taken up again by Magaard. He essentially proved the theorem in his Ph.D. thesis of 1963 [2]. This and subsequent extensions of the theorem have been discussed by Seahra and Wesson [3], who start from the Gauss-Codazzi equations and consider an alternative proof which can be applied to the induced-matter and membrane theories mentioned above.

The rediscovery of Campbell's theorem by physicists can be attributed largely to the work of Tavakol and coworkers [6]. They wrote a series of articles in the mid-1990s which showed a connection between the CM theorem and a large body of earlier results by Wesson and coworkers [4]. The latter group had been using 5D geometry as originally introduced by Kaluza and Klein to give a firm basis to an old idea of Einstein, who wished to transpose the "base-wood" of the right-hand side of his field equations into the "marble" of the left-hand side. That an effective or induced 4D energy-momentum tensor $T_{\alpha\beta}$ can be obtained from a 5D geometrical object such as the Ricci tensor $R_{AB}$ is evident from a consideration of the number of degrees of freedom involved in the problem (see the next section). The only requirement is that the 5D metric tensor be left general, and not be restricted by artificial constraints such as the "cylinder" condition imposed by Kaluza and Klein. Given a 5D line element $dS^2 = g_{AB}(x^\gamma, l) \, dx^A dx^B$ (A,B = t, xyz,l) it is then merely a question of algebra to show that the equations $R_{AB} = 0$ contain the ones $G_{\alpha\beta} = T_{\alpha\beta}$ named after Einstein [7]. Many exact solutions of $R_{AB} = 0$ are now known (see ref. 4 for a catalog). Of these, special mention should be made of the "standard" 5D cosmological ones due to Ponce de Leon [8], and the 1-body and other solutions in the "canonical" coordinates introduced by Mashhoon et al. [9]. It says something about the divide between physics and mathematics, that the connection between these solutions and the CM theorem was only made later, by the aforementioned work of Tavakol et al. [6]. Incidentally, these workers also pointed out the implications of the CM theorem for lower-dimensional ($N < 4$) gravity, which some researchers believe to be relevant to the quantization of this force.

Notwithstanding the widespread acceptance of the role of the CM theorem in physics, Anderson [5] has lately expressed a different and derogatory view of things [5]. He refers to Campbell's results as a "hodge-podge", and Magaard's results as "messy" (loc. cit., pp. 2 and 18 respectively). It is not easy to discern the scientific reasons for these opinions in his rambling discourse (which runs to 23 pages). He does, however, state that he dislikes the theorem "because it offers no guarantee of continuous dependence on the data and because it disregards causality" (loc. cit., p.1). A possible interpretation of these comments is that the CM theorem does not guarantee a well-posed initial-value problem, or the non-occurrence of singularities. However, the CM theorem does not claim to do either thing. It is a <u>local</u> embedding theorem, and cannot be pushed towards solving other problems which are the domain of (much more difficult) <u>global</u> embeddings. Quite apart from this, it is the opinion of most cosmologists that issues to do with singularities and causality are only likely to be resolved in the context of a quantum theory of gravity, something which is at present still hypothetical.

Rather, the CM theorem should be taken for what is it: a local and therefore modest result on classical embeddings, whose main "practical" implication is that we can gain a better understanding of matter in 4D by looking at the field equations in 5D.

3.  The CM Theorem and Matter

As mentioned in Section 1, there are two versions of 5D gravity which are currently popular, namely induced-matter theory and membrane theory. The Campbell-Magaard theorem is relevant to both [3], and it was shown recently by Ponce de Leon that the two theories have equivalent field equations [10]. Here, we wish to leave aside the historical considerations of Section 2, and take a fresh look at the theorem and how it relates to matter in the two noted approaches.

The CM theorem in succinct form says: Any analytic Riemannian space $V_n(s,t)$ can be locally embedded in a Ricci-flat Riemannian space $V_{n+1}(s + 1,t)$ or $V_{n+1}(s,t + 1)$.

We are here using the convention that the "small" space has dimensionality n with coordinates running 0 to n-1, while the "large" space has dimensionality n + 1 with coordinates running 0 to n. The total dimensionality is N = 1 + n, and the main physical focus is on N = 5.

The CM theorem provides a mathematical basis for the induced-matter theory, wherein matter in 4D as described by Einstein's equations $G_{\alpha\beta} = T_{\alpha\beta}$ is derived from apparent vacuum in 5D as described by the Ricci-flat equations $R_{AB} = 0$ [4,7]. The main result is that the latter set of relations satisfy the former set if

$$T_{\alpha\beta} = \frac{\Phi_{,\alpha;\beta}}{\Phi} - \frac{\varepsilon}{2\Phi^2} \left\{ \frac{\Phi_{,4} g_{\alpha\beta,4}}{\Phi} - g_{\alpha\beta,44} + g^{\lambda\mu} g_{\alpha\lambda,4} g_{\beta\mu,4} \right.$$
$$\left. - \frac{g^{\mu\nu} g_{\mu\nu,4} g_{\alpha\beta,4}}{2} + \frac{g_{\alpha\beta}}{4} \left[ g^{\mu\nu}_{,4} g_{\mu\nu,4} + \left(g^{\mu\nu} g_{\mu\nu,4}\right)^2 \right] \right\} .$$

Here the 5D line element is $dS^2 = g_{\alpha\beta}(x^\gamma, l) dx^\alpha dx^\beta + \varepsilon \Phi^2(x^\gamma, l) dl^2$, where $\varepsilon = \pm 1$, a comma denotes the ordinary partial derivative and a semicolon denotes the ordinary 4D covariant derivative. Nowadays, it is possible to prove Campbell's theorem using the ADM formalism, whose lapse-and-shift technique has been applied extensively to derive the energy of 5D solutions. It is also possible to elucidate the connection between a smooth 5D manifold (as in induced-matter theory) and one containing a singular surface (as in membrane theory). We now proceed to give an ultra-brief account of this subject.

Consider an arbitrary manifold $\Sigma_n$ in a Ricci-flat space $V_{n+1}$. The embedding can be visualized by drawing a line to represent $\Sigma_n$ in a surface, the normal vector $n^A$ to it satisfying $n \cdot n \equiv n^A n_A = \varepsilon = \pm 1$. If $e^A_{(\alpha)}$ represents an appropriate basis and the extrinsic curvature of $\Sigma_n$ is $K_{\alpha\beta}$, the ADM constraints read

$$G_{AB} n^A n^B = -\frac{1}{2}\left(\varepsilon R^\alpha_\alpha + K_{\alpha\beta} K^{\alpha\beta} - K^2\right) = 0$$

$$G_{AB} e^A_{(\alpha)} n^B = K^\beta_{\alpha;\beta} - K_{,\alpha} = 0 \ .$$

These relations provide 1 + n equations for the $2 \times n(n+1)/2$ quantities $g_{\alpha\beta}$, $K_{\alpha\beta}$. Given an arbitrary geometry $g_{\alpha\beta}$ for $\Sigma_n$, the constraints therefore form an under-determined system for $K_{\alpha\beta}$, so infinitely many embeddings are possible.

This demonstration of Campbell's theorem can easily be extended to the case where $V_{n+1}$ is a de Sitter space or anti-de Sitter space with an explicit cosmological constant, as in some applications of brane theory. Depending on the application, the remaining n(n +1) – (n + 1) = ($n^2$ – 1) degrees of freedom may be removed by imposing initial conditions on the geometry, physical conditions on the matter, or conditions on a boundary.

The last is relevant to brane theory with the $Z_2$ symmetry, where $dS^2 = g_{\alpha\beta}\left(x^\gamma, l\right) dx^\alpha dx^\beta + \varepsilon dl^2$ with $g_{\alpha\beta} = g_{\alpha\beta}\left(x^\gamma, +l\right)$ for $l \geq 0$ and $g_{\alpha\beta} = g_{\alpha\beta}\left(x^\gamma, -l\right)$ for $l \leq 0$ in the bulk. Non-gravitational fields are confined to the brane at $l = 0$, which is a singular surface. Let the energy-momentum in the brane be represented by $\delta(l) S_{AB}$ (where $S_{AB} n^A = 0$) and that in the bulk by $T_{AB}$. Then the field equations read $G_{AB} = \kappa[\delta(l)S_{AB} + T_{AB}]$ where $\kappa$ is a 5D coupling constant. The extrinsic curvature discussed above changes across the brane by an amount $\Delta_{\alpha\beta} \equiv K_{\alpha\beta}(\Sigma_{l>0}) - K_{\alpha\beta}(\Sigma_{l<0})$ which is given by the Israel junction conditions. These imply

$$\Delta_{\alpha\beta} = -\kappa\left(S_{\alpha\beta} - \frac{1}{3}Sg_{\alpha\beta}\right) \ .$$

But the $l = 0$ plane is symmetric, so

$$K_{\alpha\beta}(\Sigma_{l>0}) = -K_{\alpha\beta}(\Sigma_{l<0}) = -\frac{\kappa}{2}\left(S_{\alpha\beta} - \frac{1}{3}Sg_{\alpha\beta}\right) \ .$$

This result can be used to evaluate the 4-tensor

$$P_{\alpha\beta} \equiv K_{\alpha\beta} - Kg_{\alpha\beta} = -\frac{\kappa}{2}S_{\alpha\beta} \ .$$

However, $P_{\alpha\beta}$ is actually identical to the 4-tensor $(g_{\alpha\beta,4} - g_{\alpha\beta}g^{\mu\nu}g_{\mu\nu,4})/2\Phi$ of induced-matter theory, where it figures in 4 of the 15 field equations $R_{AB} = 0$ as $P^\beta_{\alpha;\beta} = 0$ [4,10]. That is, the conserved tensor $P_{\alpha\beta}$ of induced-matter theory is essentially the same as the total energy-momentum tensor in $Z_2$-symmetric brane theory. Other correspondences can be established in a similar fashion.

Thus while induced-matter theory and membrane theory are often presented as alternatives, they are in fact the same thing, and from the viewpoint of differential geometry both are rooted in the CM theorem.

4. A Frame for New Physics

Campbell's conjecture [1], as augmented by the work of Magaard [2], effectively provides a "ladder" to go between Riemannian manifolds whose dimensionalities differ by one. It is the basis of the now acknowledged view that the 4D Einstein equations $G_{\alpha\beta} = T_{\alpha\beta}$ are contained in the 5D Ricci equations $R_{AB} = 0$. It is difficult to count "solutions" objectively, since some are highly specific while others refer to classes; but at present, the author is aware of about 30 cases where solutions of $R_{AB} = 0$ not only contain those of $G_{\alpha\beta} = T_{\alpha\beta}$ but also provide new algebraic paths which may have physical significance [3,4]. The chance of this occurring in the absence of Campbell's theorem are less than $(1/2)^{30}$, which is infinitesimal. Thus, while detractors may wish for ever-more rigorous proofs of the theorem [5], it is accepted by the community as a foundation.

What, then, are its implications for new physics? Specifically, what physics (different from that already known for N = 4) might one expect by an application of Campbell's theorem to N-dimensional relativity?

Lower-dimensional gravity (N = 2,3) is somewhat trivial, but is studied by some pedestrians who use the motivation that such formalisms lend themselves more readily to quantization than the N = 4 version due to Einstein, which we know to be an excellent (if non-quantized) description of Nature. However, these wanderers in the lower-dimensional desert frequently invent their field equations, sheltering behind the well-known pathologies of Riemannian geometry in 2 and 3 dimensions. (For example, in 3 dimensions the Riemann-Christoffel tensor may be written as a function of the Ricci tensor alone.) If there is any academic advantage to studying N = (2,3) gravity, the field equations ought at least to conform to what one obtains by using the Campbell ladder to descend from N = 4 gravity.

Higher-dimensional gravity $(N \geq 5)$ is an algebraically rich field, but one which requires great physical insight to yield meaningful results. With one extra dimension, the extra variable treated covariantly can turn up in numerous guises; and of course the number of independent field equations increases from 10 (as in Einstein theory) to 15 (as in Kaluza-Klein theory and its modern variants). The trick here is to tie down the extra variable in a physical way by using the five coordinate degrees of freedom to put the 5D line element (metric) into a form which relates unambiguously to the 4D one. It is this trick which underlies the wealth of results which flow from the "canonical" coordinates (see above). Metrics of this form show that the extra coordinate is related to the concept of particle (rest) mass in classical physics, or alternatively the quantum measure of mass which arises from the Higgs field of particle physics. Consider, now, the implications of the 5D Ricci-flat field equations $R_{AB} = 0$. Solutions of these have been known for many years which show that the big-bang singularities of the 4D Friedmann-Robertson-Walker models are merely the result of a bad choice of coordinates in a smooth, Riemann-flat 5D manifold. (The coordinate transformations necessary to show this are complicated but known: see ref. 4, p. 49 for the transformations to a space with $R_{ABCD} = 0$, and ref. 8 for the original solutions which reduce to the standard 4D cosmological ones on hypersurfaces of metrics with $R_{AB} = 0$.) While not all solutions that are Ricci-flat ($R_{AB} = 0$) are also Riemann-flat ($R_{ABCD} = 0$) one can argue that both kinds of manifold are in some sense free of matter or energy. The implication of this for particles is clear: they should travel on <u>null</u> 5D geodesics. This idea has recently been taken up in the literature, and has a considerable future. It means that what we perceive as massive particles in 4D are akin to photons in 5D.

For N > 5, the Campbell ladder will be of even greater significance. The number of independent components of the Ricci tensor in ND is $N(N + 1)/2$. Going up from N = 4 (Einstein theory) through N = 5 (Kaluza-Klein, or induced-matter and membrane theory), one comes to certain values of N which have special properties. For example, the Schwarzschild solution can only be embedded in a flat space of $N \geq 6$. Further, any solution of the ND Einstein equations can be embedded in a flat space if $N \geq 10$. (This is basically because in 4D there are 10 independent components of the metric tensor, so in 10D they can all be set to constants, a property which in particle physics is related to supersymmetry, where every integral-spin boson is paired with a half-integral spin fermion, to produce a state with zero energy.) For certain values of N, it is possible to relate the classical properties of a manifold to the quantum properties of particles in a convenient manner, which has led to special interest in N = 11 and N = 26. (This is basically because the internal symmetry groups of particles need to be accommodated, and because the divergent self-energies of point particles can be avoided if they are regarded as tiny strings.) However, from the perspective of differential geometry, no value of N is sacrosanct. The value of N needs to be chosen with a view to the physics to be explained. But whatever the value of N, Campbell's ladder will have to be considered. Stepping off it will lead, as regards physical interpretation, into a morass.

5.  Conclusion

Campbell's theorem was outlined in 1926, which coincidentally is the year when Klein added a quantum twist to the 5D version of classical relativity due to Kaluza, a topic which was endorsed by and worked on later by Einstein. One can only speculate on what course physics might have taken had Campbell's conjecture been widely known in those years. As things developed, it was not until 1963 that the theorem was proven with some rigor by Magaard, and not until the 1990s that its importance for physics was recognized by Tavakol and others. It is now a central theme of studies in higher-dimensional relativity. It shows that Einstein's 4D equations $G_{\alpha\beta} = T_{\alpha\beta}$ are embedded in the Ricci-flat 5D equations $R_{AB} = 0$, providing thereby a rationale for induced-matter and membrane theory. It also provides a ladder by which one can ascend to even higher dimensions, as in supersymmetry, supergravity and string theory.

Riemannian geometry was shown by Einstein to provide an excellent formalism for describing gravitational physics in 4D spacetime. Campbell's theorem shows how to extend that formalism to other dimensions, and is expected to play a basic role in any ND theory of physics.

Acknowledgements

Thanks for comments go to S.S. Seahra and R. Tavakol. Thanks for support go to N.S.E.R.C.

Note Added

After completing the above, two important papers appeared which have direct relevance to the argument [11, 12]. Katzourakis has used bundle theory to relax the locality assumption in Campbell's theorem, and shown that any pseudo-Riemannian analytic n-manifold can be embedded isometrically in an (n + 1)D bulk with arbitrary topology, provided the Ricci and torsion tensors vanish [11]. He has concluded that any solution of the 4D Einstein equations can be embedded in a Ricci-flat manifold with one or more extra dimensions. This is basically what Tavakol and Wesson with their coworkers have been doing for years to study physics. Aguilar and Bellini [12] have used this embedding theorem to study the extra force and particle mass in a 5D theory, from the global viewpoint. Their results agree with similar studies by Wesson and coworkers [4], from the local viewpoint.

References

1. Campbell, J.E., 1926. A Course of Differential Geometry. Clarendon Press, Oxford.

2. Magaard, L., 1963. Ph.D. Thesis, Kiel.

3. Seahra, S.S., Wesson, P.S., 2003. Class. Quant. Grav. $\underline{20}$, 1321.

4. Wesson, P.S., 1999. Space, Time, Matter. World Scientific, Singapore.

5. Anderson, E., 2004. gr-gc/0409122.

6. Rippl, S., Romero, C., Tavakol, R., 1995. Class. Quant. Grav. $\underline{12}$, 2411.

   Romero, C., Tavakol, R., Zalaletdinov, R., 1996. Gen. Rel. Grav. $\underline{28}$, 365.

   Lidsey, J.E., Romero, C., Tavakol, R., Rippl, S., 1997. Class. Quant. Grav. $\underline{14}$, 865.

7. Wesson, P.S., Ponce de Leon, J., 1992. J. Math. Phys. $\underline{33}$, 3883.

8. Ponce de Leon, J., 1988. Gen. Rel. Grav. $\underline{20}$, 539.

9. Mashhoon, B., Liu, H., Wesson, P.S., 1994. Phys. Lett. B $\underline{331}$, 305.

10. Ponce de Leon, J., 2001. Mod. Phys. Lett. A $\underline{16}$, 2291.

    Ponce de Leon, J., 2002. Int. J. Mod. Phys. D$\underline{11}$, 1355.

11. Katzourakis, N.I., 2004. Math-ph/0407067.

12. Aguilar, J.E.M., Bellini, M., 2004. qr-gc/0408054.